\documentclass[conference]{IEEEtran}
\usepackage{amsmath,amssymb,amsfonts}
\usepackage{algorithmic}
\usepackage{graphicx}
\usepackage{textcomp}
\usepackage[table,xcdraw]{xcolor}
\usepackage{booktabs}
\usepackage{rotating}
\usepackage{tikz}

\begin{document}

\title{A Novel Epidemiological Approach to Geographically Mapping Population Dry Eye Disease in the United States through Google Trends\\
}

\author{\IEEEauthorblockN{Daniel B. Azzam B.S.}
\IEEEauthorblockA{\textit{School of Medicine} \\
\textit{University of California, Irvine}\\
Irvine, USA \\
azzamd@hs.uci.edu}
\and
\IEEEauthorblockN{Nitish Nag, Ph.D.}
\IEEEauthorblockA{\textit{Donald Bren School of Information and Computer Sciences} \\
\textit{University of California, Irvine}\\
Irvine, USA \\
nagn@uci.edu}
\and
\IEEEauthorblockN{Julia Tran, B.S.}
\IEEEauthorblockA{\textit{School of Medicine} \\
\textit{University of California, Irvine}\\
Irvine, USA \\
}
\and
\IEEEauthorblockN{Lauren Chen, B.S.}
\IEEEauthorblockA{\textit{School of Medicine} \\
\textit{University of California, Irvine}\\
Irvine, USA \\
}
\and
\IEEEauthorblockN{Kaajal Visnagra B.S.}
\IEEEauthorblockA{\textit{School of Medicine} \\
\textit{University of California, Irvine}\\
Irvine, USA \\
}
\and
\IEEEauthorblockN{Kailey Marshall, O.D.}
\IEEEauthorblockA{\textit{School of Medicine} \\
\textit{University of California, Irvine}\\
Irvine, USA \\
}
\and
\IEEEauthorblockN{Matthew Wade, M.D.}
\IEEEauthorblockA{\textit{School of Medicine} \\
\textit{University of California, Irvine}\\
Irvine, USA \\
wadem@hs.uci.edu}
}





%

\maketitle

\begin{abstract}
Dry eye disease (DED) affects approximately half of the United States population. DED is characterized by dryness on the corena surface due to a variety of causes. This study fills the spatiotemporal gaps in DED epidemiology by using Google Trends as a novel epidemiological tool for geographically mapping DED in relation to environmental risk factors. We utilized Google Trends to extract DED-related queries estimating user intent from 2004-2019 in the United States. We incorporated national climate data to generate heat maps comparing geographic, temporal, and environmental relationships of DED. Multi-variable regression models were constructed to generate quadratic forecasts predicting DED and control searches. Our results illustrated the upward trend, seasonal pattern, environmental influence, and spatial relationship of DED search volume across US geography. Localized patches of DED interest were visualized along the coastline. There was no significant difference in DED queries across US census regions. Regression model 1 predicted DED searches over time (R\textsuperscript{2}=0.97) with significant predictors being control queries (p=0.0024), time (p=0.001), and seasonality (Winter p=0.0028; Spring p<0.001; Summer p=0.018). Regression model 2 predicted DED queries per state (R\textsuperscript{2}=0.49) with significant predictors being temperature (p=0.0003) and coastal zone (p=0.025). Importantly, temperature, coastal status, and seasonality were stronger risk factors of DED searches than humidity, sunshine, pollution, or region as clinical literature may suggest. Our work paves the way for future exploration of geographic information systems for locating DED and other diseases via online search query metrics.
\end{abstract}

\begin{IEEEkeywords}
population health, Google Trends, healthy lifestyle, personalized health, precision health.
\end{IEEEkeywords}

\section{Introduction}
Google Inc. first initiated Google Trends in 2006 as a source of data that compares the popularity of search terms on Google Search in various regions and languages from around the world \cite{Alphabet2006GoogleTrends,Jun2018TenApplications}. Google Trends is a tool that is free, readily available worldwide, and updated daily at https://Google.com/trends/.
In 2009, the Centers for Disease Control and Prevention (CDC) partnered with Google and developed the first ever use of data from internet search engines to successfully predict flu outbreaks in advance \cite{Ginsberg2009DetectingData}. To this day, Google Trends has successfully predicted trends in the utbreaks of many viruses, including West Nile virus, norovirus, varicella, influenza, and HIV \cite{Ginsberg2009DetectingData,Johnson2014ATrends,Jena2013PredictingData,Pelat2009MoreTrends,Wilson2009EarlyInternet}. More recently, Google Trends has been used to study an increasing variety of healthcare domains, such as diabetes \cite{Nuti2014TheReview,Tkachenko2017GoogleDiabetes}. 
Dry eye disease is historically one of the most commonly seen conditions in the ophthalmology clinic, with a significant detrimental impact on patients’ quality of life \cite{Gayton2009EtiologyDisease,OBrien2004DryStrategies}. Epidemiological data for DED has traditionally been collected via surveys requiring excessive time and resources, while providing only limited data specific to the time and population that is studied, such as the Women’s Health Study, the Physicians’ Health Studies, and the 2013 National Health and Wellness Survey \cite{Schaumberg2003PrevalenceWomen,Schaumberg2009PrevalenceStudies,Farrand2017PrevalenceOlder}. The 2017 Tear Film and Ocular Surface Society Dry Eye Workshop II (TFOS DEWS II) reported that in the last 10 years there have been no population studies examining DED prevalence in any region of the world south of the equator \cite{Craig2017TFOSSummary,Andras2018NewII}.
In this study, we provide the most current estimates of dry eye disease prevalence across the United States using Google Trends data. Our analysis fills the gaps in current dry eye disease epidemiology, which was largely derived from surveys, by reaching beyond the limitations of time and space. We provide an analysis of how internet search intent can be extrapolated to map DED geographically and how this can be compared in relation to environmental risk factors such as temperature and humidity. This work serves as a proof of concept of using internet epidemiological tools along with GIS data from the environment as a mapping technique for disease surveillance models. This GIS approach can be universally applied to population diseases across the globe. Clinically, Google search analysis of dry eye search intent may one day be able to power ophthalmology clinic logistics according to real time changes in dry eye risk factors such as climate or season.

\section{MATERIALS AND METHODS}
The experiments were done using public domain data and analysis tools. These are described below.

\subsection{Google Trends}
The following methodologies were designed based on published methods \cite{Stein2013GaugingSurgery}. The Google Trends search interest output numbers are relative to the peak point of popularity for the most popular search term within the query in the given region and time \cite{Alphabet2006GoogleTrends}. The numbers range from 0 to 100, with a value of 100 meaning peak popularity and a value of 50 meaning that the term is half as popular at that time point. These values were then normalized relative to control search terms for the given region and time and are represented as arbitrary units (a.u.) from 0 a.u. to 100 a.u.
On August 5, 2019, we completed a series of queries in Google Trends with query filters confined to within the region of the United States over the time period of January 1, 2004 to August 1, 2019 under all categories within the web search Google property. We downloaded data for a list of DED symptoms and treatments that are frequently encountered in clinical practice: “dry eyes”, “irritated eyes”, “scratchy eyes”, “watery eyes”, “burning eyes”, “gritty eyes”, “eye drops”, “artificial tears”, “Restasis”, “Systane”, “Oasis tears”, “Thera tears”, “Sooth”, “Blink tears”, “Visine”, “Clear eyes”, “Xiidra”, and “punctal plugs”. The scale from 0 to 100 a.u. was maintained the same by including the most popular search term, “eye drops”, in each query.
Control search terms included: “news”, “weather”, “sports”, and “Google”. We summed the search interest of all the DED terms together at each time point and all the control terms together at each time point, then normalized DED search interest relative to control search interest at each time point on a scale of 0 to 100 a.u.

\subsection{Environmental Factors}
We downloaded climate and historical weather data from the United States National Climatic Data Center \cite{NationalOceanicandAtmosphericAdministration2019DataNCDC}. This data included annual state temperature average, average afternoon humidity, and average annual sunshine from 1971 to 2000. Air pollution rankings were obtained from U.S. News and World Report \cite{USNews2020StateStates}. Air quality data were gathered from the U.S. Environmental Protection Agency \cite{EPA2019AirUS}. Wind speed data was collected from www.USA.com \cite{WorldMediaGroup2020U.S.Rank}.

\subsection{Statistical Analysis}
The values for normalized DED search intent relative to control search intent, temperature, and humidity were categorized by United States (US) census regions based on the US Census Bureau. Results between US census regions of West, Midwest, South, and Northeast were analyzed for differences using unpaired Brown-Forsythe One-Way ANOVA (GraphPad Prism 5.0; GraphPad Software, CA, USA). Statistical significance was determined at p<0.05. The Brown-Forsythe ANOVA was applied here because the standard deviations were not equal between regions. We generated geographic heat maps in order to compare the geographic, temporal, and environmental relationships of dry eye disease queries. Additionally, we conducted linear regression analysis to compare the relationship between DED search queries with temperature, and separately, with humidity. For predictive analytics, multi-variable regression models were constructed to generate quadratic forecasts to predict dry eye search intent and control search intent.

\section{RESULTS}
In the following section we describe various relationships between user search query trends and factors that are understood to cause dry eye disease.

\subsection{Temporospatial Trends in DED Search Intent in Relation to Temperature and Humidity}
\begin{figure*}[htbp]
\centerline{\includegraphics[scale=0.5]{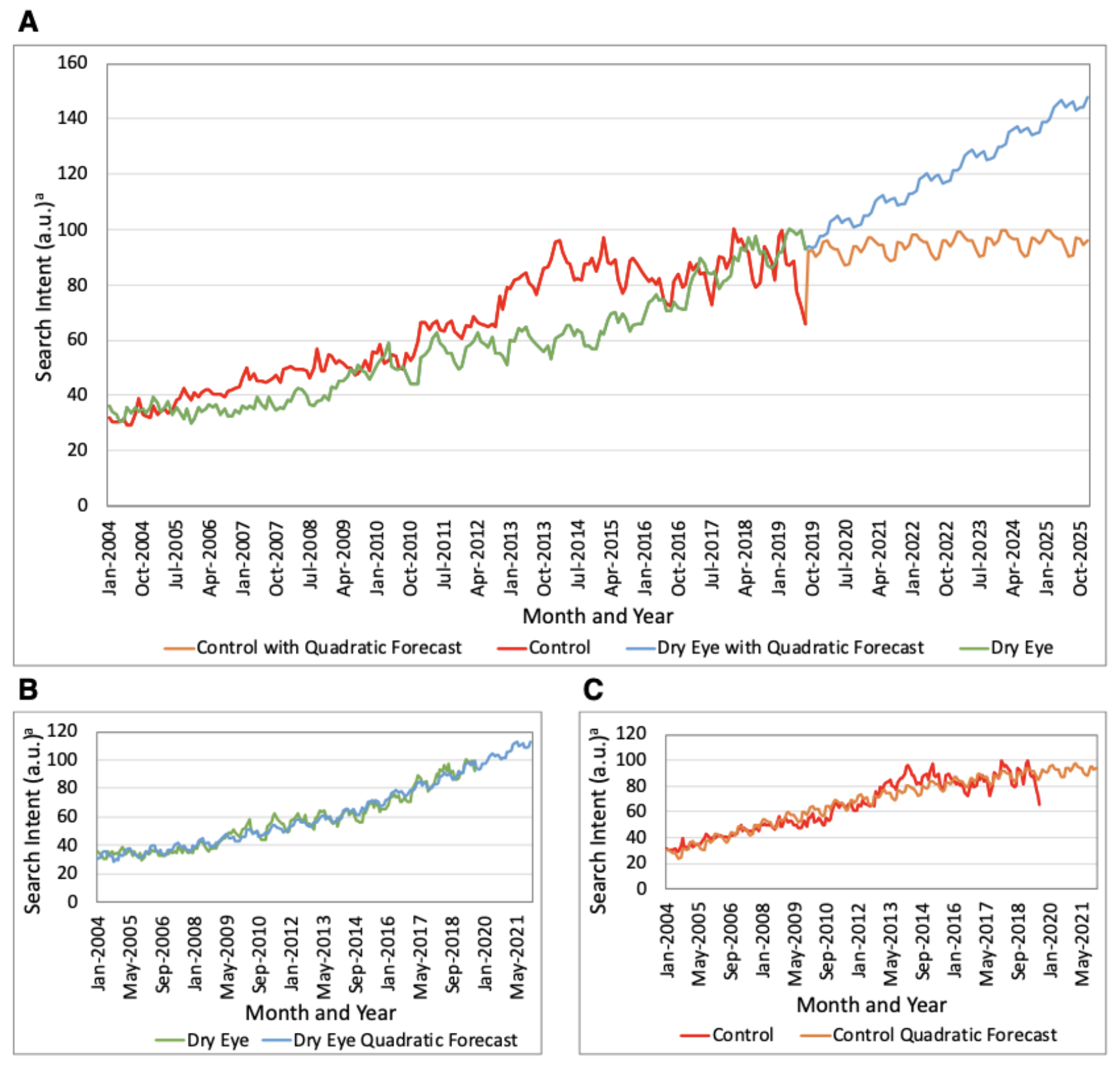}}
\caption{\textbf{Time series plot of Search Intent for dry eye and control terms in the United States from 2004 to 2019 with forecasts to 2025:} The data are generated from Google Trends data for search queries related to dry eye disease and control queries. Quadratic forecasts were generated using multi-variable regression models, as shown in Table \ref{tab:2}, to predict dry eye search intent and control search intent. A: Time series plot illustrating dry eye and control search intent from 2004 to 2019 with forecasts to 2025. There is a significant upward trend and seasonal pattern in dry eye search intent over time. B: Time series plot illustrating dry eye search intent overlaid with the dry eye quadratic forecast. C: Time series plot illustrating control search intent overlaid with the control quadratic forecast in arbitrary units (a.u.) where Google sets the highest search intent during 2004-2019 to 100 a.u.}
\label{fig:1}
\end{figure*}

The Google Trends data from January 2004 to August 2019 illustrated a significant upward trend such that dry eye search intent grew 157\%, while control search intent grew 106\%, with a significant pattern of seasonality [Figure \ref{fig:1}]. Looking forward, dry eye search intent is forecasted to grow an additional 57\% to 145.47 a.u. (95\% CI, 138.30,152.63 a.u.) by July 2025 and 441\% to 500.76 a.u. (95\% CI, 493.60, 507.92 a.u.) by July 2040.

\begin{figure*}[htbp]
\centerline{\includegraphics[scale=0.30]{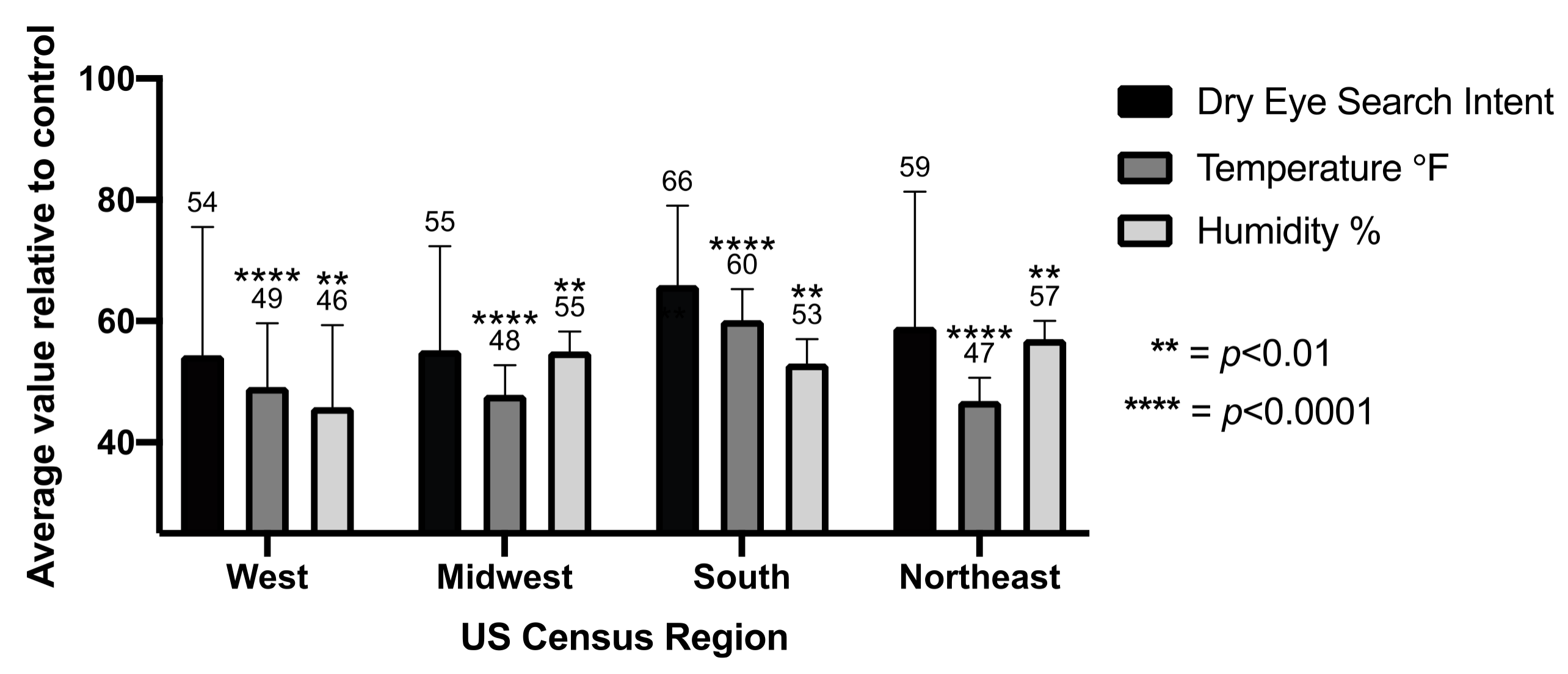}}
\caption{\textbf{Bar graph for comparisons of dry eye Google search queries, temperature, and humidity by United States census region:} The data for dry eye search queries, temperature, and humidity were averaged across the four US census regions. These sets of data were separately compared in Brown-Forsythe ANOVA statistical analysis. p-values less than 0.05 were considered significant, with ** denoting p<0.01, and **** denoting p<0.0001.}
\label{fig:2}
\end{figure*}

The data for DED search intent relative to control search intent, temperature, and humidity were graphed by US census region [Figure \ref{fig:2}]. There was no significant difference in dry search queries across the regions of West, Midwest, South, and Northeast (Brown-Forsythe One-Way ANOVA, p=0.35). However, temperature and humidity were significantly different across census regions (Brown-Forsythe One-Way ANOVA, p<0.0001 and <0.01, respectively), with the Northeast being the coldest (47°F) and the most humid (57\%) region. The hottest region was the South (60°F), while the driest region was the West (46\%).

\begin{figure*}[htbp]
\centerline{\includegraphics[scale=0.35]{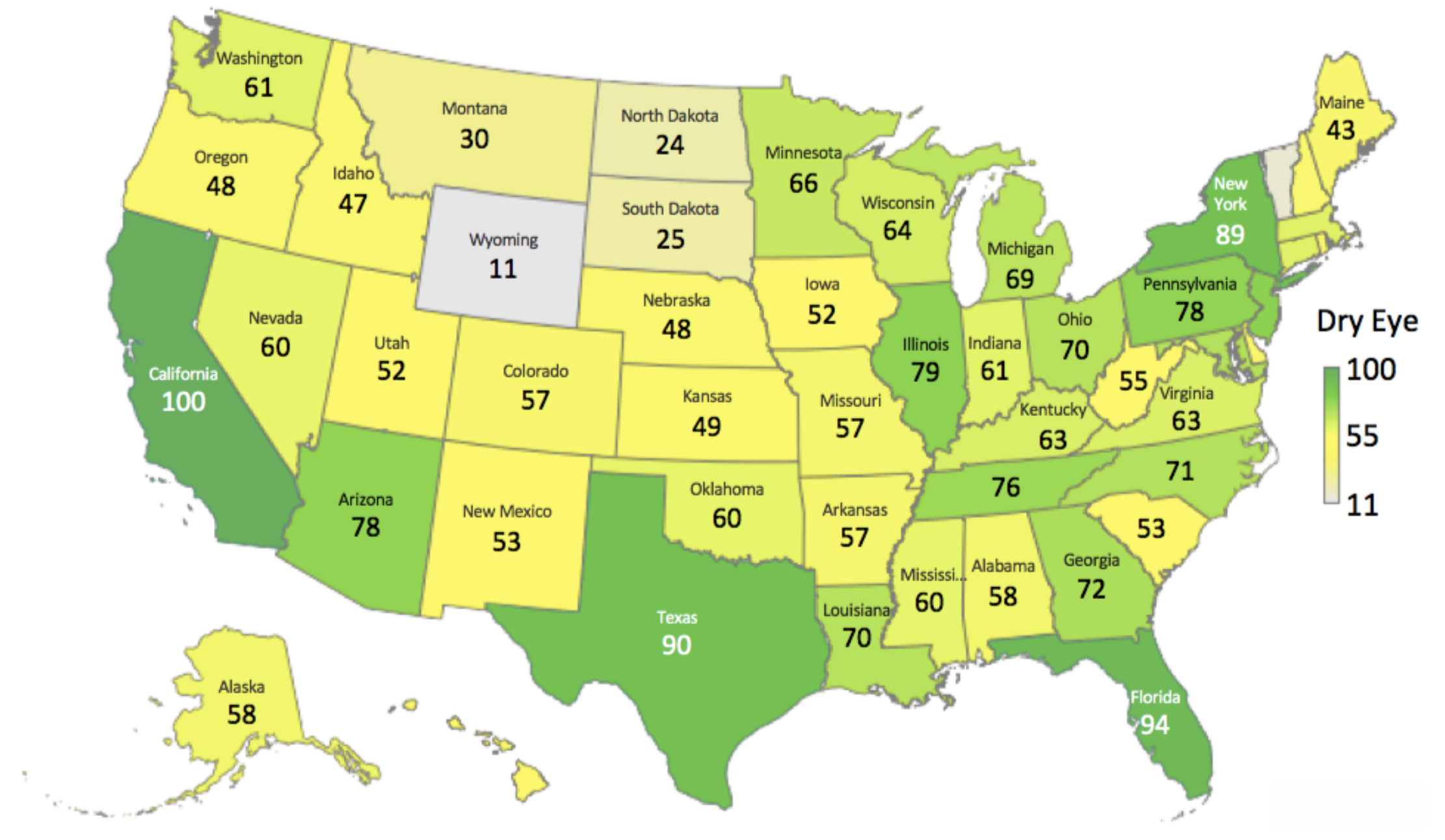}}
\caption{\textbf{Geographic heat map of dry eye search intent on Google Trends in the United States:} This map was generated with the data from Google Trends relative to the control search terms. Localized patches of interest can easily be visualized in various hot spots across the country, with the majority of these hot spots being along the coast.}
\label{fig:3}
\end{figure*}

The spatial relationship of dry eye search queries was mapped across the United States geography, illustrating the highest relative dry eye search intent in California and the least in Wyoming [Figure \ref{fig:3}].

\begin{figure*}[htbp]
\centerline{\includegraphics[scale=0.35]{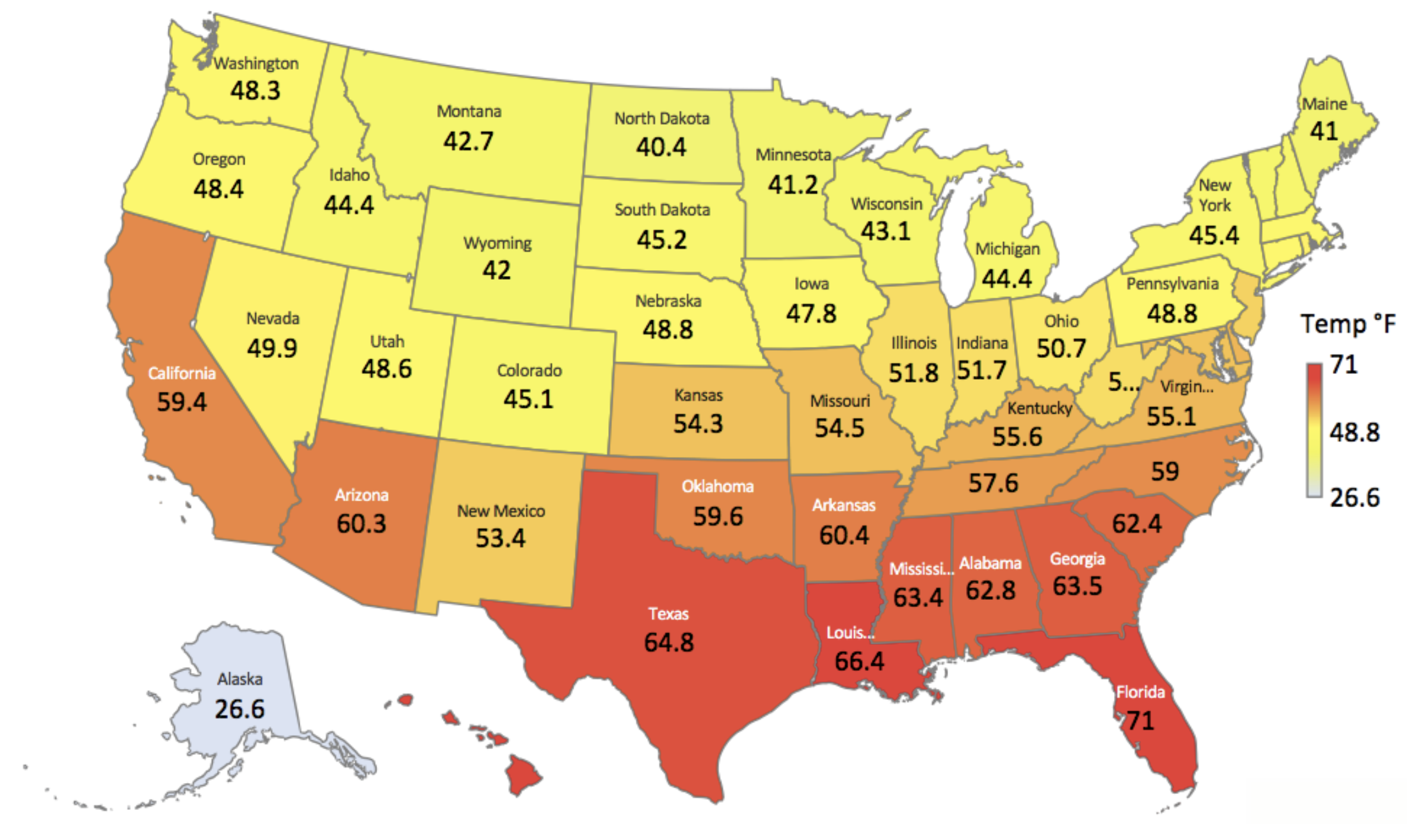}}
\caption{\textbf{Geographic heat map of average annual temperatures (°F) in the United States:} This map was generated using data from the United States National Climatic Data Center.}
\label{fig:4}
\end{figure*}

Localized hot spots for interest in dry eye exist, mostly along the coastline. The average temperature was mapped across the United States, with the highest temperature in Florida (70.7°F) and the lowest in Alaska (26.6°F) [Figure \ref{fig:4}].

\begin{figure*}[htbp]
\centerline{\includegraphics[scale=0.35]{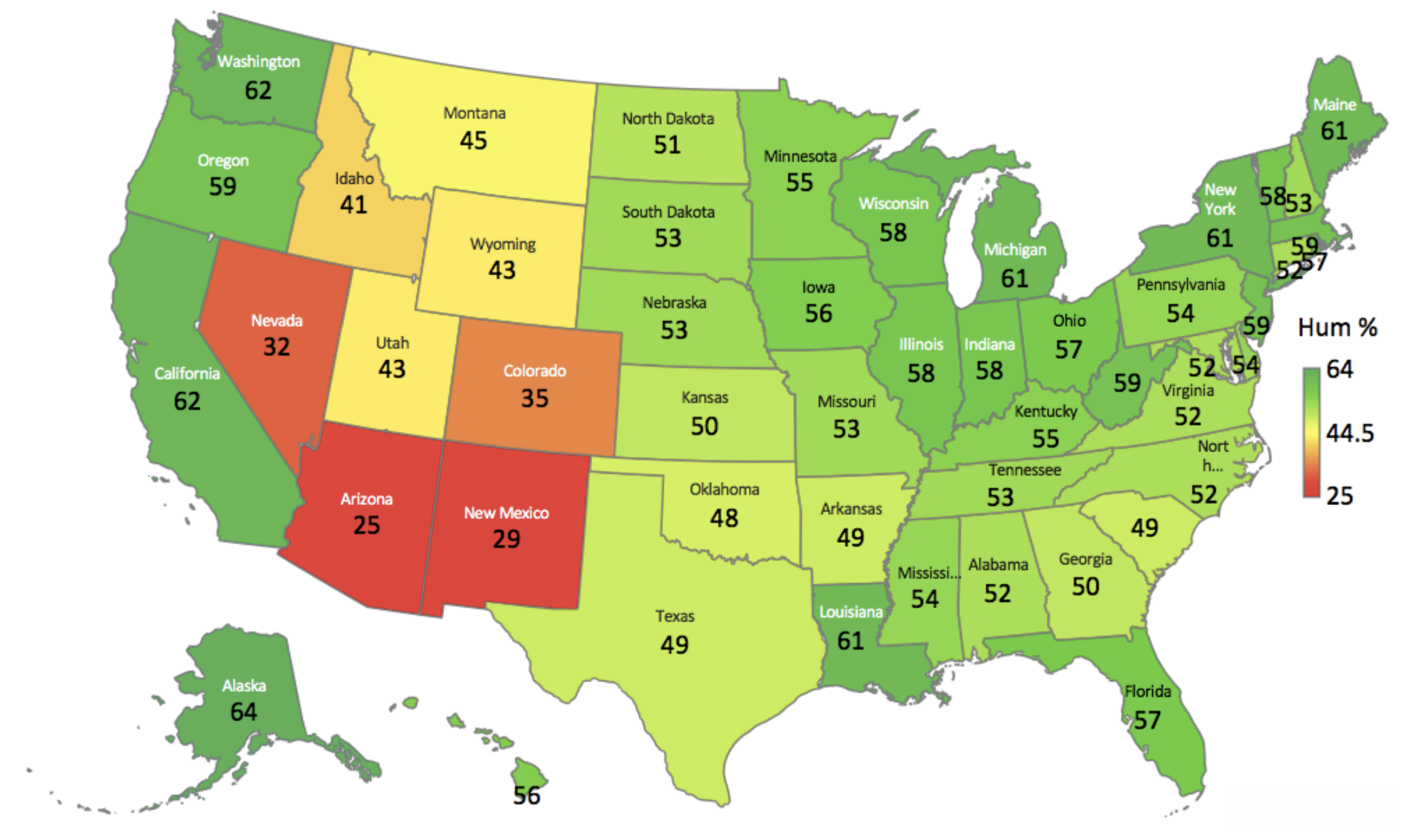}}
\caption{\textbf{Geographic heat map of average afternoon humidity (\%) in the United States:} This map was generated using data from the United States National Climatic Data Center.}
\label{fig:5}
\end{figure*}

The average afternoon humidity was mapped across the United states, with the highest humidity in Alaska (64\%) and the lowest in Arizona (25\%) [Figure \ref{fig:5}].

\begin{figure}[htbp]
\centerline{\includegraphics[scale=0.45]{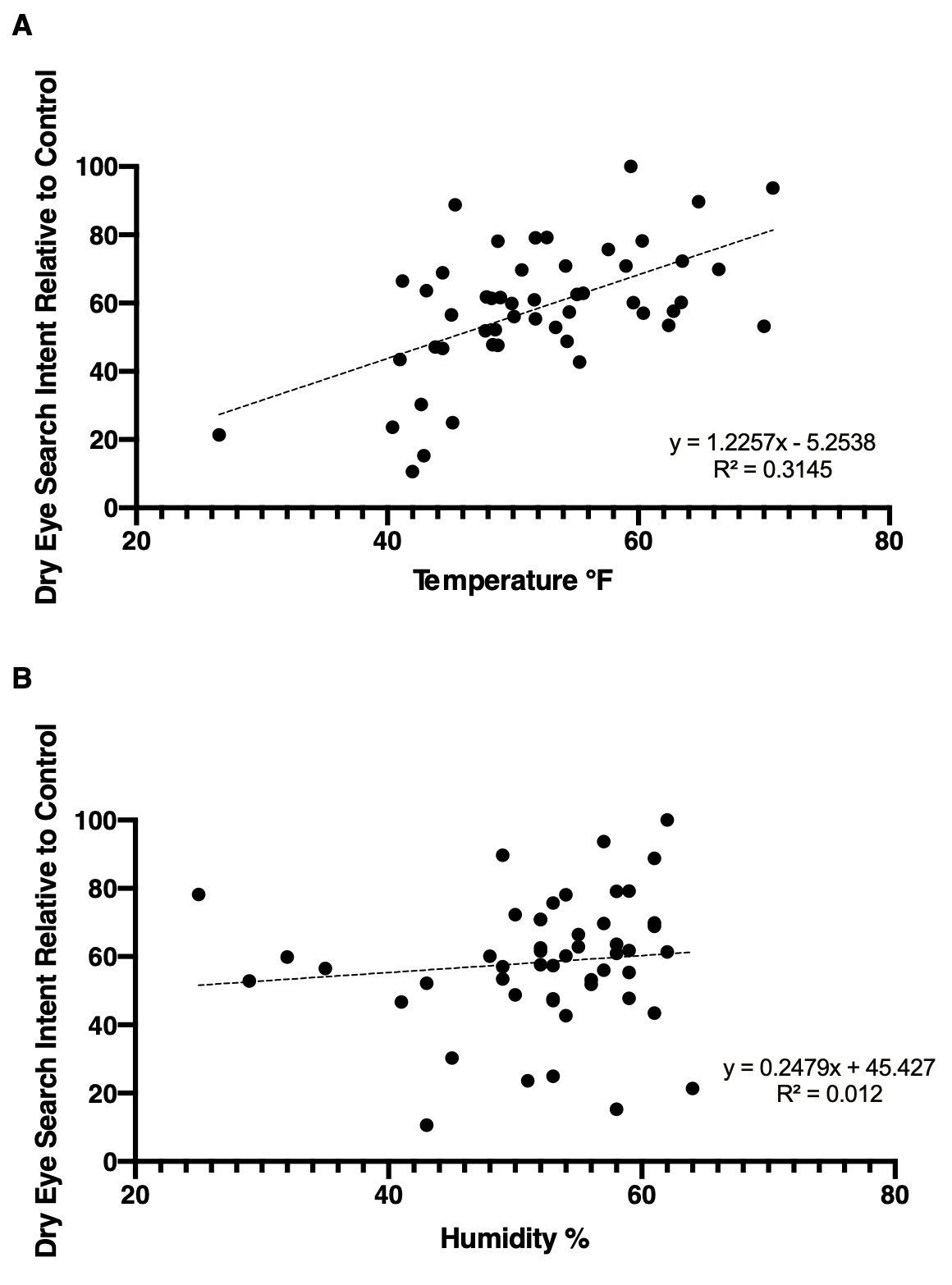}}
\caption{\textbf{Scatter plot for comparison of Google Trends data for dry eye and temperature or humidity in the United States:} A linear model was used to analyze the data, with the equation of the line of best fit and the coefficient of determination, R2, listed. A: Dry eye search intent versus temperature. B: Dry eye search intent versus humidity.}
\label{fig:6}
\end{figure}

\begin{figure}[htbp]
\centerline{\includegraphics[scale=0.69]{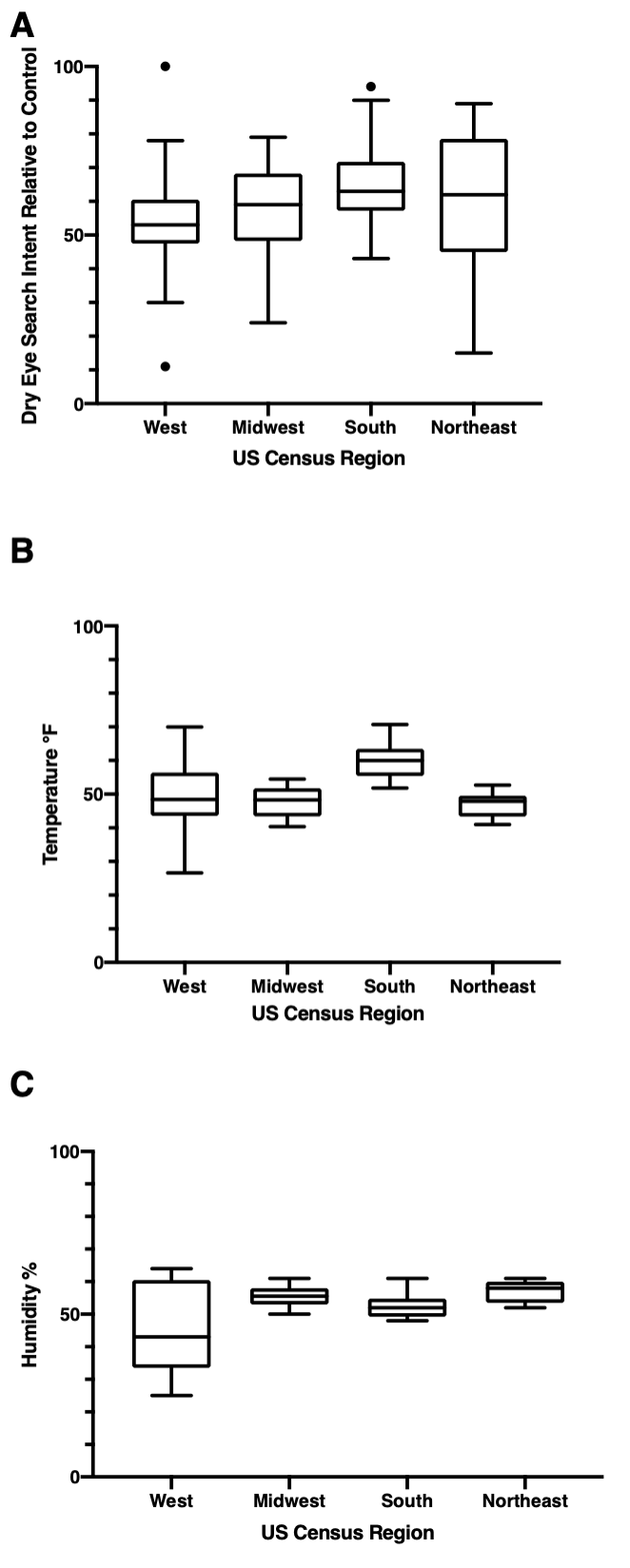}}
\caption{\textbf{Box and whisker diagrams for comparison of (A) dry eye search intent, (B) temperature, and (C) humidity across US census regions:} A: There were 2 outliers for dry eye search intent in the West, California (100 a.u.) and Wyoming (11 a.u.), and 1 outlier in the South, Florida (94 a.u.). B, C: There were no outliers for temperature or humidity.}
\label{fig:7}
\end{figure}

Temperature was compared to dry eye search queries in Figure \ref{fig:6}, which showed a positive linear correlation with moderate strength (r = 0.56). Humidity was compared to dry eye search queries in Figure \ref{fig:6}, which illustrated no linear relationship (r = 0.11). The distribution of the data for dry eye search intent, temperature, and humidity across the US regions was plotted in Figure \ref{fig:7}. There were 3 outlier states for dry eye search intent, including California (100 a.u.) and Wyoming (11 a.u.) in the West and Florida (94 a.u.) in the South Figure \ref{fig:7}. There were no outlier states for temperature or humidity.

\subsection{Multi-variable Regression Models and Predictive Analytics}
Table \ref{tab:1} demonstrates the multi-variable regression model using state-specific data such as geography and environmental risk factors to predict DED search intent relative to control per state, and Table \ref{tab:3} shows the corresponding correlation coefficients for all variables that were considered. Variables that were significant risk factors of DED search intent were temperature (r = 0.56, p<0.001) and coastal zone (r = 0.43, p=0.025). Using the model in Table \ref{tab:1}, there is a 1.32 a.u. (95\% CI, 0.64, 1.99 a.u.) increase in DED search intent with each 1 degree Fahrenheit increase in temperature, and DED search intent in coastal states is 13.28 a.u. (95\% CI, 1.74, 24.82 a.u.) higher than non-coastal states. No relationship exists between DED search intent and humidity (r = 0.11, p=0.48) or US census regions (West r = -0.21, p=0.45; South r = 0.28, p=0.15; Northeast r = 0.02, p=0.64).

\begin{table}[]
\centering
\resizebox{.48\textwidth}{!}{%
\begin{tabular}{@{}lrrrr@{}}
\toprule
\textbf{Variables} &
  \multicolumn{1}{l}{\textbf{Coefficients}} &
  \multicolumn{1}{l}{\textbf{Standard error}} &
  \multicolumn{1}{l}{\textbf{t-Statistic}} &
  \multicolumn{1}{l}{\textbf{P-value}} \\ \midrule
Intercept             & -7.12  & 27.31 & -0.26 & 0.796 \\
Temperature           & 1.32   & 0.34  & 3.93  & 0.001 \\
Humidity              & -0.25  & 0.35  & -0.71 & 0.479 \\
Rank in Low Pollution & 0.25   & 0.17  & 1.46  & 0.152 \\
Coastal               & 13.28  & 5.72  & 2.32  & 0.025 \\
West region           & -5.05  & 6.57  & -0.77 & 0.447 \\
South region          & -10.24 & 6.91  & -1.48 & 0.146 \\
Northeast region      & 3.41   & 7.21  & 0.47  & 0.639 \\ \midrule
\textbf{R\textasciicircum{}2 = 0.49} &
  \multicolumn{1}{l}{} &
  \multicolumn{1}{l}{} &
  \multicolumn{1}{l}{} &
  \multicolumn{1}{l}{}
\end{tabular}%
}
\caption{Multi-variable regression model using environmental risk factors and eye search intent relative to control. This model predicts average dry eye search intent relative to control for each state based on that state’s average temperature, humidity, air pollution, coastal zone, and US census region. The regions are relative to the Midwest.}
\label{tab:1}
\end{table}

\begin{table}[]
\centering
\resizebox{.35\textwidth}{!}{%
\begin{tabular}{@{}lr@{}}
\toprule
\textbf{Variables}    & \multicolumn{1}{l}{\textbf{Correlation coefficient, r}} \\ \midrule
Temperature           & 0.56                                                    \\
Humidity              & 0.11                                                    \\
\% Sun                & 0.10                                                    \\
Sunshine hours        & 0.06                                                    \\
Wind speed            & -0.46                                                   \\
Rank in low pollution & 0.31                                                    \\
Coastal               & 0.43                                                    \\
Atlantic coast        & 0.21                                                    \\
Pacific coast         & -0.03                                                   \\
Gulf coast            & 0.28                                                    \\
Great Lakes           & 0.30                                                    \\
West region           & -0.21                                                   \\
Midwest region        & -0.09                                                   \\
South region          & 0.27                                                    \\
Northeast             & 0.02                                                    \\ \midrule
                      & \multicolumn{1}{l}{}                                   
\end{tabular}%
}
\caption{Correlation coefficients relating dry eye search intent relative to control with various variables for each state.}
\label{tab:3}
\end{table}

Table \ref{tab:2} demonstrates the multi-variable regression model using time, season, and environmental risk factors to predict DED search intent over time, and Table \ref{tab:4} shows the corresponding correlation coefficients for all variables that were considered. Variables that were significant predictors of DED search intent were control search intent (r = 0.85, p=0.0024), time (r = 0.96, p<0.001), time2 (r = 0.97, p<0.001), and seasonality (Winter r = -0.04, p=0.0028; Spring r = 0.10, p<0.001; Summer r = 0.05, p=0.018). Using the model in Table \ref{tab:2}, there is a 0.16 a.u. (95\% CI, 0.097, 0.22 a.u.) increase each month in DED search intent; DED search intent in Winter, Spring, and Summer are 3.68 a.u. (95\% CI, 1.28, 6.07 a.u.), 6.47 a.u. (95\% CI, 4.96, 7.98 a.u.), and 2.83 a.u. (95\% CI, 0.49, 5.17 a.u.) higher than in Fall, respectively.

\begin{table}[]
\centering
\resizebox{.48\textwidth}{!}{%
\begin{tabular}{@{}lrrrr@{}}
\toprule
\rowcolor[HTML]{FFFFFF} 
\textbf{Variables} &
  \multicolumn{1}{l}{\cellcolor[HTML]{FFFFFF}\textbf{Coefficients}} &
  \multicolumn{1}{l}{\cellcolor[HTML]{FFFFFF}\textbf{Standard error}} &
  \multicolumn{1}{l}{\cellcolor[HTML]{FFFFFF}\textbf{t-Statistic}} &
  \multicolumn{1}{l}{\cellcolor[HTML]{FFFFFF}\textbf{P-value}} \\ \midrule
\rowcolor[HTML]{FFFFFF} 
Intercept                & 31.11 & 2.85 & 10.93 & 0.001 \\
\rowcolor[HTML]{FFFFFF} 
Control Searches         & -0.14 & 0.04 & -3.08 & 0.002 \\
\rowcolor[HTML]{FFFFFF} 
Temperature              & 0.06  & 0.05 & 1.18  & 0.24  \\
\rowcolor[HTML]{FFFFFF} 
AQI value                & -0.02 & 0.02 & -0.84 & 0.403 \\
\rowcolor[HTML]{FFFFFF} 
Time\textasciicircum{}2  & 0.00  & 0    & 10.75 & 0.001 \\
\rowcolor[HTML]{FFFFFF} 
Time                     & 0.16  & 0.03 & 5.04  & 0.001 \\
\rowcolor[HTML]{FFFFFF} 
Winter                   & 3.68  & 1.21 & 3.03  & 0.003 \\
\rowcolor[HTML]{FFFFFF} 
Spring                   & 6.47  & 0.76 & 8.47  & 0.001 \\
\rowcolor[HTML]{FFFFFF} 
\textit{\textbf{Summer}} & 2.83  & 1.19 & 2.38  & 0.018 \\ \midrule
\textbf{R\textasciicircum{}2 = 0.97} &
  \multicolumn{1}{l}{} &
  \multicolumn{1}{l}{} &
  \multicolumn{1}{l}{} &
  \multicolumn{1}{l}{}
\end{tabular}%
}
\caption{Multi-variable regression model using environmental risk factors, time (month and year), and season to predict dry eye search intent. This model predicts dry eye search intent for each month based on control search intent, temperature, AQI value, time, and season of the year. Seasons are relative to Fall.}
\label{tab:2}
\end{table}

\begin{table}[]
\centering
\resizebox{.35\textwidth}{!}{%
\begin{tabular}{@{}lr@{}}
\toprule
\textbf{Variables} & \multicolumn{1}{l}{\textbf{Correlation coefficient, r}} \\ \midrule
Control Searches   & 0.85                                                    \\
Temperature        & 0.06                                                    \\
AQI value          & -0.11                                                   \\
Time2              & 0.97                                                    \\
Time               & 0.96                                                    \\
Winter             & -0.04                                                   \\
Spring             & 0.10                                                    \\
Summer             & 0.05                                                    \\
January            & 0.00                                                    \\
February           & 0.00                                                    \\
March              & 0.03                                                    \\
April              & 0.06                                                    \\
May                & 0.06                                                    \\
June               & 0.03                                                    \\
July               & 0.02                                                    \\
August             & 0.02                                                    \\
September          & -0.05                                                   \\
November           & -0.06                                                   \\
December           & -0.06                                                   \\ \midrule
                   & \multicolumn{1}{l}{}                                   
\end{tabular}%
}
\caption{Correlation coefficients relating dry eye search intent relative to control with various variables for each state.}
\label{tab:4}
\end{table}

\section{DISCUSSION}
To our knowledge, this is the first study to use internet search intent to geographically map dry eye disease in comparison to the unique regional characteristics of the United States by using statistical analysis of national Google Trends and environmental data. Traditional prevalence data for dry eye disease has been collected via surveys, such as the 2013 National Health and Wellness Survey, that require an excessive amount of time and resources to conduct \cite{Schaumberg2003PrevalenceWomen,Farrand2017PrevalenceOlder,TearFilmandOcularSurfaceSociety2007Introduction2007,Schein1999DryAssessment,Schein1997PrevalenceElderly,Munoz2000CausesStudy,Moss2000PrevalanceSyndrome,Christen1998Low-dosePhysicians,Christen2000DesignTrials,Castro2018PrevalenceQuestionnaire}. Additionally, these types of surveys are limited in space and time to the participants examined in a certain geographic region within a certain year and are laden with subjective biases and observer effects. According to the 2007 Dry Eye Workshop (DEWS) report, we face a recurring issue that there is still a need to conduct epidemiological studies for dry eye in different geographical populations \cite{TearFilmandOcularSurfaceSociety2007Introduction2007}. The most recent TFOS DEWS II report published in 2017, stated that there have been zero population studies examining DED prevalence in the southern hemisphere in the last 10 years, with most of the DED prevalence studies focusing the attention on Asia and Europe \cite{Craig2017TFOSSummary,Andras2018NewII}.
The solution to this scarcity of epidemiological data may be digital epidemiology: the use of real time monitoring of disease through internet data, which overcomes the issues of resources, time, and geographic region. In the world of today, this digital data method is becoming a reality. Internet data can be used to instantly track data from all over the world over any period of time to include all people who have access to internet. Furthermore, the internet data is accessible to researchers at no additional cost.
An example of the successful use of internet data for epidemiology can be seen in the field of infectious diseases. In 2009, the CDC partnered with researchers from Google in order to develop the first ever use of data from internet search engines to predict infectious disease trends \cite{Ginsberg2009DetectingData,Johnson2014ATrends}. Using terms for flu-like symptoms, the CDC and Google were capable of successfully predicting flu outbreaks weeks in advance of the CDC’s US Influenza Sentinel Provider Surveillance Network. After this project made international headlines, Google made its program, “Google Trends,” publicly available. To this day, Google Trends has successfully predicted trends in the outbreaks of many viruses, including West Nile virus, norovirus, varicella, influenza, and HIV \cite{Ginsberg2009DetectingData,Johnson2014ATrends,Jena2013PredictingData,Pelat2009MoreTrends,Wilson2009EarlyInternet}.
First described in 1952, the topic of dry eye disease has gained significant popularity. DED is one of the most frequent ophthalmic morbidities in the United States, with 25\% of patients in ophthalmology clinics complaining of dry eye \cite{Gayton2009EtiologyDisease,OBrien2004DryStrategies}. Recent estimates show that about 30 million people suffer from dry eye disease in the US \cite{Farid2017DryBox,Paulsen2014DryLife}. The tremendous incidence of DED signifies the increasing importance of monitoring and treating patients who suffer from dry eyes.
Dry eye disease is a chronic multifactorial ocular surface disease in which the surface of the eye is inadequately lubricated due to pathophysiology involving a continuum of evaporative dry eye and poor tear quantity or quality \cite{Farrand2017PrevalenceOlder,Craig2017TFOSSummary,TearFilmandOcularSurfaceSociety2007Introduction2007,NationalEyeInstitute2020HyperopiaInstitute}[14,15,22,32]. Risk factors include old age, female sex, environment (low humidity, extremes of temperature, air pollution), cigarette smoking, screen time or blink rate, refractive surgery, contact lens use, several medications (such as anti-depressants), and several conditions (auto-immune disease, inflammatory disease, and aqueous or mucinous abnormalities) \cite{Gayton2009EtiologyDisease}. The presenting symptoms of DED vary from a burning sensation to blurry vision, ocular pain, or the sensation of a foreign body in the eye \cite{NationalEyeInstitute2020HyperopiaInstitute}. Treatments include ocular lubricants (such as artificial tears), various anti-inflammatory medications, punctal plugs, modification of local environment, dietary modifications (including oral essential fatty acid supplementation), lid hygiene, and warm compresses \cite{Galor2015DryParameters,Craig2017TFOSSummary}. Despite these therapies, patients often will continue to have several DED symptoms.
Dry eye disease plays a major socioeconomic impact on our society. In the US, DED patients visit the doctor an average of 6 times per year, costing a total of \$800 USD; this totals to a national cost of \$4 billion USD per year \cite{Drew2018ReflectionsProducts,Yu2011TheAnalysis}. Furthermore, considering the loss of productivity,estimations of the annual financial burden of DED in the United States exceed \$55 billion USD. The true costs of DED on society are greater when considering the effects of DED on quality of life, vision, and productivity, as well as the psychological and physical impact of pain \cite{Craig2017TFOSSummary}. This overall socioeconomic burden motivates the need to better identify geographic regions with DED and relieve dry eye symptoms at a low cost. This study aims to address this problem by examining the potential to use of readily available internet data, such as Google Trends, to geographically map DED-related search intent in the US.
Google search intent for dry eye disease in the US population has grown by 157\% from 2004 to 2019. The dry eye disease “epidemic” will continue to expand, with forecasts predicting that DED search intent will grow an additional 57\% by 2025 and 441\% by 2040. This rise reflects the epidemiological trends reported from national survey data, with an increase in US DED prevalence from 4.34\% in 2009 to 6.8\% in 2017 \cite{Schaumberg2009PrevalenceStudies,Farrand2017PrevalenceOlder}. This demonstrates the concept that as the prevalence of overall DED rises in the United states, so does the volume of dry eye information-seeking behavior on Google Trends.
Furthermore, dry eye search intent showed a significant seasonal pattern with the highest search volume in Spring and lowest in Fall, which confirms previous findings by Kumar et al. in 2015 \cite{Kumar2015SeasonalEye}. When compared across the four US census regions, our results demonstrated no significant difference in DED search intent across the regions, which matches traditional reports of dry eye regional prevalence in 2017 \cite{Farrand2017PrevalenceOlder}. Nevertheless, regional differences in dry eye prevalence have been observed in various other countries, including Korea, Taiwan, and India \cite{Jie2009PrevalenceStudy,Han2011PrevalencePopulation,Wang2012ComorbiditiesStudy,Sahai2005DryPopulation}. Thus, it comes as a surprise that there is no regional DED prevalence differences in the US despite vast differences in climate and environment. The United States is comprised of a wide variety of natural ecosystems, ranging from desserts to wetlands \cite{NationalGeographic2020Asia:Society,RutledgeK.RamroopT.BoudreauT.McDanielM.SantaniT.SproutE.CostaH.Hun2011EcosystemSociety}. Perhaps many of these ecosystems are present in all 4 US census regions, and therefore the grouping itself may be hiding unique geographic differences in DED search intent. When looking at individual states, localized patches of hot spots exist, such as California and Florida, mostly along the coastline. This may be explained by the high speed coastal winds, as high wind speed has been shown to be significantly correlated with dry eye disease \cite{NationalWeatherService2020CoastalWinds,Um2014SpatialKorea}. Understanding the unique nature of these regions and how they play a role on the geographic trends in DED adds an important perspective to the investigation of DED in this nation.
Regarding environmental factors, our results demonstrated that temperature and coastal status were more important predictors of DED search intent than humidity, sunshine duration, air pollution, or geographic region. A study conducted in South Korea found that meteorological factors such as higher temperature, lower humidity, higher wind speed, longer sunshine duration, and higher air pollution were all significantly correlated with DED \cite{Um2014SpatialKorea}. Human studies have demonstrated that humidity was associated with measures such as tear evaporative rate, blink rate, and tear volume \cite{Um2014SpatialKorea,Uchiyama2007IncreasedSyndrome,Teson2013InfluenceDisease,Lopez-Miguel2014DryConditions,Alex2013FactorsStress}. In the current study, the relationship between DED and humidity may be drowned out by other stronger relationships. Furthermore, the observed geospatial patterns of DED might be due to regional differences in a myriad of factors spanning beyond just the environmental components studied. These factors might include regional differences in lifestyle including screen habits, demographics (age, sex, ethnicity, etc.), socioeconomic factors, awareness, advertising and media exposure, or various other types of noise e.g. internet bots or users tampering with Google to alter market size for financial reasons. Further investigation is needed for the association between population DED search intent and the various underlying factors contributing to the trends.
While mining the web for epidemiology in real time is a fascinating perspective, in the present day this type of internet data should not be assumed to replace any efforts of public health organizations or clinicians \cite{Cervellin2017IsSettings}. As we move toward the future of reporting internet epidemiologic data in health care, greater transparency in the methodology through which we gather our internet data can improve its reliability as a research tool and bring us closer to real time digital epidemiology \cite{Nuti2014TheReview}.
On the topic of dry eye disease, future studies should expand our understanding of DED populations by examining the relationship between DED search intent and various other regional factors e.g. meteorological, demographic, socioeconomic, lifestyle. This data in the US can then be compared to trends across the world. In the future, hospitals may be able to monitor diseases like DED in real time based on the unique regional characteristics of their own community via publicly available internet data. This data can then be used to power hospital logistics according to the changing incidence of a limitless variety of illnesses in different months of the year. Beyond the study of disease at the macro level with population digital epidemiology, future work should investigate diseases, such as dry eye disease, at the micro level for individual patients. With lifestyle pattern tracking in today’s world, patients can collect personal data including daily habits, climate, geospatial data, and screen time \cite{Reeves2019Screenomics:Them,Nag2017HealthObservations,Pandey2018UbiquitousHealth,Pandey2020ContinuousRetrieval}. In the future, this type of data can one day be incorporated to monitor and advise on an individual’s personal eye health state \cite{Nag2018Cross-modalEstimation,Nag2020HealthEstimation,Nag2019SynchronizingMonitoring}. This individual monitoring allows for personal health navigation to control the health state of the eye to the best potential for a certain individual \cite{Nag2019ALife,Nag2017CyberneticHealth,Nag2020HealthEstimation}.
Several limitations are present in our study. First, interpretation of population trends in US DED via Google Trends is challenging without the clinical information provided by traditional surveys such as medical comorbidities and symptom severity. Second, there is a possibility of internet data being influenced by various unknown factors, like media exposure, or altered by users or bots. Third, the data sets comparing DED search intent to temperature and humidity were consolidated to average data-points per state based on the climate data available, limiting the sample size. At the same time, this sample of fifty states may have been skewed by outliers e.g. California or Wyoming. Additionally, the search terms utilized might reflect other conditions of the ocular surface e.g. allergic conjunctivitis, rather than being specific to DED. On the same note, Google users may have entered synonyms of dry eye symptoms or treatments that were missed in our collection.

\section{CONCLUSIONS}
Our study is the first to demonstrate that internet search intent such as Google Trends can be used as a novel digital epidemiologic approach for geographically mapping population dry eye disease in relation to the unique regional characteristics of the United States. This type of mapping allows for easy visualization of localized hot spots in dry eye search intent, which were mostly located along the coastline. The interest in dry eye in the United States grew tremendously since 2004, with an overall upward trend and a seasonal variation pattern. Meteorological factors varied in their relation to dry eye, with temperature and coastal status being more important predictors of DED search intent than humidity, sunshine duration, air pollution, or geographic region. This paper creates an avenue for the future exploration of geographic information systems for locating dry eye and other diseases through online population disease metrics. Further investigation is needed to determine what other regional factors are contributing to the differences in DED search intent across the nation, how these results compare to dry eye trends across the world, and how this information can be applied at the micro level to monitor an individual’s personal eye health state \cite{Nag2018Cross-modalEstimation,Nag2020HealthEstimation}. Continuous access to estimating this personal eye health state allows for the foundation by which individual's may recieve automated guidance on how to best ensure their corneal health stays well \cite{Nag2019ALife}.

\bibliographystyle{plain}
\bibliography{references}

\end{document}